\documentclass[]{revtex4}  %>>> use for US letter paper
\input{Qcircuit}
\newcommand{\dbra}[1]{{\langle\langle{#1}\vert}}
\newcommand{\dket}[1]{{\vert{#1}\rangle\rangle}}
\usepackage{latexsym,amsmath,amssymb,amsbsy,graphicx}
\usepackage[utf8]{inputenc}
\begin{document}  
  
  \title{Measure of decoherence in quantum error correction for solid-state quantum computing} 
  
  \author{Alexey A. Melnikov}
  \email{melnikov@phystech.edu}
    
  \author{Leonid E. Fedichkin}%
   \email{leonidf@gmail.com}
    \altaffiliation[Also at ]{NIX, Moscow, Russia}%Lines break automatically or can be forced with \\
    \affiliation{%
    Institute of Physics and Technology, Russian Academy of Sciences, Moscow, Russia
    }%
    \affiliation{%
    Moscow Institute of Physics and Technology, Dolgoprudny, Russia
    }%

%%%%%%%%%%%%%%%%%%%%%%%%%%%%%%%%%%%%%%%%%%%%%%%%%%%%%%%%%%%%% 
\begin{abstract}
We considered the interaction of semiconductor quantum register with noisy environment leading to various types of qubit errors. We analysed both phase and amplitude decays during the process of electron-phonon interaction. The performance of quantum error correction codes (QECC) which will be inevitably used in full scale quantum information processors was studied in realistic conditions in semiconductor nanostructures. As a hardware basis for quantum bit we chose the quantum spatial states of single electron in semiconductor coupled double quantum dot system. The modified 5- and 9-qubit quantum error correction (QEC) algorithms by Shor and DiVincenzo without error syndrome extraction were applied to quantum register. 5-qubit error correction procedures were implemented for Si charge double dot qubits in the presence of acoustic phonon environment. $ \chi $-matrix, Choi–Jamiołkowski state and measure of decoherence techniques were used to quantify qubit fault-tolerance. Our results showed that the introduction of above quantum error correction techniques at small phonon noise levels provided quadratic improvement of output error rates. The efficiency of 5-qubits quantum error correction algorithm in semiconductor quantum information processors was demonstrated.
\end{abstract}

 \maketitle

%>>>> Include a list of keywords after the abstract 

%\keywords{Quantum error correction, Si double dot qubit, quantum channel, measure of decoherence, chi-matrix, Choi–Jamiołkowski state}

%%%%%%%%%%%%%%%%%%%%%%%%%%%%%%%%%%%%%%%%%%%%%%%%%%%%%%%%%%%%%
\section{INTRODUCTION}
\label{sec:intro}  % \label{} allows reference to this section
In recent years much attention is attracted to the influence of decoherence on quantum communication channels~\cite{holevo2010quantum, holevo2011probabilistic, nielsen2010quantum} and quantum information processing~\cite{nielsen2010quantum, preskill1998lecture, fedichkin2005study, fedichkin2009quantitative}. Measure of decoherence approach~\cite{fedichkin2003measures, fedichkin2004additivity} occurred to be helpful tool for quantitative evaluation of quantum state distortion due to noisy environment. $ \chi $-matrix and Choi–Jamiołkowski state representations are very efficient for quantum channel description and study~\cite{sudarshan1961stochastic, holevo1972mathematical, choi1973positive, holevo2010quantum, holevo2011probabilistic, nielsen2010quantum, chruscinski2009spectral}. In this paper we combine these approaches to analyse the influence of error correction~\cite{shor1995scheme, divincenzo1996fault, steane1996simple, steane1996error, caves1999quantum, tomita2011unitary, li2012recovery} on quantum registers. The representative example of solid-state qubit -- electron in semiconductor double dot interacting with phonons is also considered.

%%%%%%%%%%%%%%%%%%%%%%%%%%%%%%%%%%%%%%%%%%%%%%%%%%%%%%%%%%%%%
\section{DESCRIBING QUANTUM CHANNELS}
We consider a qubit system described by a density matrix $\rho$ of size $ d\times d  $. By definition, the operator $\rho$ is Hermitian, positive semidefinite and has trace one~\cite{fano1957description, blum2012density, holevo2010quantum, holevo2011probabilistic, nielsen2010quantum}. An evolution of a density matrix is described by a quantum operation $\mathcal{E}$ (also called a quantum channel or a stochastic map)~\cite{sudarshan1961stochastic, holevo1972mathematical, choi1973positive, holevo2010quantum, holevo2011probabilistic, nielsen2010quantum, chruscinski2009spectral}. We use these terms interchangeably. As usual a linear map $\mathcal{E}$ is called a quantum operation when it preserves trace (in Schrödinger picture) and is completely positive~\cite{holevo2010quantum}. In our investigation, a qubit system is placed in an environment that acts on these qubits independently. That means that the quantum operation describing the impact of the environment on the system of qubits is separable. Choi has shown~\cite{choi1975completely} that every quantum channel can be represented as a sum of Kraus operators (Kraus representation~\cite{kraus1983states}):
\begin{equation}\label{eq:kraus_summ}
\mathcal{E}[\rho]=\sum_{i=1}^n \hat{E}_i \rho \hat{E}^\dagger_i,\qquad \sum_{i=1}^n \hat{E}^\dagger_i \hat{E}_i=\hat{I}.
\end{equation}
This statement is also a sufficient condition for a map to be a quantum channel. Equation~(\ref{eq:kraus_summ}) shows a way to describe a channel.

There are several ways to describe a quantum channel mathematically. In the sake of convenience, we will use a Choi-Jamiołkowski state~\cite{jamiolkowski1972linear} $ \tau $ in simulation of quantum circuits and a $\chi$-matrix representation in the calculation of measure of decoherence. Below we consider both of these representations.

%%-----------------------------------------------------------
\subsection{$\chi$-matrix representation}

Every linear map $\mathcal{E}$ which is a quantum operation can be written in the form of:
\begin{equation}\label{eq:operator_summ}
\mathcal{E}[\rho]=\sum_{\substack{\alpha,\beta=0}}^{d^2-1}\chi_{\alpha\beta}\hat{E}_{\alpha}\rho \hat{E}_{\beta}^\dag,
\end{equation}
where the set of $\{\hat{E}_{\alpha}\}_{\alpha=0}^{d^2-1}$ matrices is the basis in $\mathcal{H}_d^{\otimes 2}$ space.
\medskip

The coefficients $\chi_{\alpha\beta}$ form $\chi$-matrix of dimension $d^2\times d^2$~\cite{nielsen2010quantum}. From the construction of $\chi$ it is seen that the matrix is Hermitian. Since $ \left(\mathcal{E}[\rho]\right)^{\dag}=\mathcal{E}[\rho]$, we have
\begin{equation}\label{eq:chi_ermit}
\left(\sum_{\substack{\alpha,\beta=0}}^{d^2-1}\chi_{\alpha\beta}\hat{E}_{\alpha}\rho \hat{E}_{\beta}^\dag\right)^\dag=\sum_{\substack{\alpha,\beta=0}}^{d^2-1}\chi_{\alpha\beta}^*\hat{E}_{\beta}\rho \hat{E}_{\alpha}^\dag=\sum_{\substack{\alpha,\beta=0}}^{d^2-1}\chi_{\alpha\beta}\hat{E}_{\alpha}\rho \hat{E}_{\beta}^\dag=\sum_{\substack{\alpha,\beta=0}}^{d^2-1}\chi_{\beta\alpha}\hat{E}_{\beta}\rho \hat{E}_{\alpha}^\dag.~\mathrm{Thus}~\chi_{\beta\alpha}=\chi_{\alpha\beta}^*.
\end{equation}

Another limitation is imposed by the fact that channel $ \mathcal{E} $ preserves trace because it is a quantum operation. Since $ \mathrm{Tr}\left[\mathcal{E}\rho\right]=\mathrm{Tr}\left[\sum_{\substack{\alpha,\beta=0}}^{d^2-1}\chi_{\alpha\beta}\hat{E}_{\alpha}\rho \hat{E}_{\beta}^\dag\right]=\sum_{\substack{\alpha,\beta=0}}^{d^2-1}\mathrm{Tr}\left[\chi_{\alpha\beta}\rho \hat{E}_{\beta}^\dag\hat{E}_{\alpha}\right]=\mathrm{Tr}\left[\rho\sum_{\substack{\alpha,\beta=0}}^{d^2-1} \chi_{\alpha\beta}\hat{E}_{\beta}^\dag\hat{E}_{\alpha}\right]=\mathrm{Tr}[\rho] $, we obtain
\begin{equation}\label{eq:chi_trace}
\sum_{\substack{\alpha,\beta=0}}^{d^2-1} \chi_{\alpha\beta}\hat{E}_{\beta}^\dag\hat{E}_{\alpha}=\hat{I}.
\end{equation}

Since $\chi$-matrix of $ 4\times4 $ dimension is Hermitian, it can be defined by 16 real numbers. The limitations due to trace conservation shown in Eq.~(\ref{eq:chi_trace}) allows us to describe $\chi$ with 12 real parameters. $\chi$-matrix formalism provides a simple method of verification of complete positivity. A map $\mathcal{E}$ is completely positive if and only if $\chi$-matrix is positive semidefinite~\cite{nambu2005matrix}.

%%-----------------------------------------------------------
\subsection{Choi-Jamiołkowski state $ \tau $ representation}

Operator $ \hat{\hat{\chi}} $ is given by
\begin{equation}\label{eq:choi_jam}
\hat{\hat{\chi}}=\sum_{\substack{\alpha,\beta=0}}^{d^2-1}\chi_{\alpha\beta}\dket{\hat{E}_{\alpha}}\dbra{\hat{E}_{\beta}},
\end{equation}
where $\dket{\hat{A}}\equiv [A_{11},...,A_{1d},A_{21},...,A_{2d},...,A_{d1},...,A_{dd}]^T$ is a supervector in Liouville space (L-space).
\medskip Supervectors $\{\dket{\hat{E}_{\alpha}}\}_{\alpha=0}^{d^2-1}$ form the basis in $\mathcal{H}_d^{\otimes 2}$. \medskip

Choi-Jamiołkowski state is defined as
\begin{equation}\label{eq:tau_jam}
\tau=(\mathcal{E}\otimes \hat{I})\ket{\Omega_{d^2}}\bra{\Omega_{d^2}},
\end{equation}
where the state $\ket{\Omega_{d^2}}=\frac{1}{\sqrt{d}}\sum_{\substack{i=0}}^{d-1}\ket{i}\ket{i}$ is a maximally entangled state. It is worth noting that $\frac{1}{\sqrt{d}}\dket{\hat{I}}=\ket{\Omega_{d^2}}$ and $\dket{\hat{A}}=\left(\hat{A}\otimes\hat{I}\right)\dket{\hat{I}}$. These properties are followed from the above definitions. The quantum operation is uniquely determined by the Choi-Jamiołkowski state. This fact is known as Choi-Jamiołkowski isomorphism. Let us show that $ \tau=\hat{\hat{\chi}}/d $:

\begin{equation}\label{eq:tau_equal_chi}
\hat{\hat{\chi}}=\sum_{\substack{\alpha,\beta=0}}^{d^2-1}\chi_{\alpha\beta}\left(\hat{E}_{\alpha}\otimes\hat{I}\right)\dket{\hat{I}}\dbra{\hat{I}}\left(\hat{E}_{\beta}\otimes\hat{I}\right)^\dag= \left(\mathcal{E}\otimes \hat{I}\right)\dket{\hat{I}}\dbra{\hat{I}}= d\left(\mathcal{E}\otimes \hat{I}\right)\ket{\Omega_{d^2}}\bra{\Omega_{d^2}} =\tau d.
\end{equation}

%%%%%%%%%%%%%%%%%%%%%%%%%%%%%%%%%%%%%%%%%%%%%%%%%%%%%%%%%%%%%
\section{MEASURE OF DECOHERENCE}

To estimate the quality of quantum correction we use the concept of measure of decoherence $D$~\cite{fedichkin2003measures, fedichkin2004additivity, fedorov2004evaluation, fedichkin2009quantitative}. By the definition, measure of decoherence is the maximum over all the states of the operator norm of matrix $\rho_{\rm{out}}-\rho_{\rm{in}} $:

\begin{equation}\label{eq:measureDdefinition}
D=\sup_{\substack{\rho_{\rm{in}}}}\left|\left|\rho_{\rm{out}}-\rho_{\rm{in}}\right|\right|,
\end{equation}
where the operator norm of Hermitian matrix $ A $ is given by $ \left|\left|A\right|\right|=\max_{\substack{a\in \mathrm{spec}(A)}}\left|a\right| $~\cite{holevo2010quantum}. By $ \mathrm{spec}(A) $ denote the spectrum of operator $ A $.

Let us consider a qubit ($ d=2 $). In this article it is shown that if the influence of the environment has the Kraus representation in Pauli basis with $ \chi $-matrix in the form of
\begin{equation}\label{eq:chi_diag}
\chi=\begin{pmatrix} \chi_0 & 0 & 0 & 0\\ 0 & \chi_1 & 0 & 0\\ 0 & 0 & \chi_2 & 0\\ 0 & 0 & 0 & \chi_3\end{pmatrix},
\end{equation}
then the expression for measure of decoherence is simplified
\begin{equation}\label{eq:D_diag_chi}
D=\max\{\chi_1+\chi_2,\chi_1+\chi_3,\chi_2+\chi_3\}=\chi_1+\chi_2+\chi_3-\min\{\chi_1,\chi_2,\chi_3\}.
\end{equation}

Let $ \chi_{mk}=\chi^{\rm{re}}_{mk}+i\chi^{\rm{im}}_{mk}  $, then using Eq.~\ref{eq:chi_ermit} we get the relations $ \chi^{\rm{re}}_{mk}=\chi^{\rm{re}}_{km}, \chi^{\rm{im}}_{mk}=-\chi^{\rm{im}}_{km} $. Using Eq.~\ref{eq:chi_trace} we obtain a system of 4 equations:
\begin{equation}
\left\{
\begin{array}{rcl}
\chi_{00}&=&1-\chi_{11}-\chi_{22}-\chi_{33}\\
\chi^{\rm{im}}_{12}&=&\chi^{\rm{re}}_{30}\\
\chi^{\rm{im}}_{31}&=&\chi^{\rm{re}}_{20}\\
\chi^{\rm{im}}_{23}&=&\chi^{\rm{re}}_{10}.\\
\end{array}
\right.
\end{equation}
Consider an arbitrary $ \chi $-matrix of $ 4\times4 $ dimension:
\begin{equation}
\chi=
\begin{pmatrix} 
1-\chi^{\rm{re}}_{11}-\chi^{\rm{re}}_{22}-\chi^{\rm{re}}_{33} & \chi^{\rm{re}}_{01}+i\chi^{\rm{im}}_{01} & \chi^{\rm{re}}_{02}+i\chi^{\rm{im}}_{02} & \chi^{\rm{re}}_{03}+i\chi^{\rm{im}}_{03}\\ \chi^{\rm{re}}_{01}-i\chi^{\rm{im}}_{01} & \chi^{\rm{re}}_{11} & \chi^{\rm{re}}_{12}+i\chi^{\rm{re}}_{03} & \chi^{\rm{re}}_{13}-i\chi^{\rm{re}}_{02}\\
\chi^{\rm{re}}_{02}-i\chi^{\rm{im}}_{02} & \chi^{\rm{re}}_{12}-i\chi^{\rm{re}}_{03} & \chi^{\rm{re}}_{22} & \chi^{\rm{re}}_{23}+i\chi^{\rm{re}}_{01}\\
\chi^{\rm{re}}_{03}-i\chi^{\rm{im}}_{03} & \chi^{\rm{re}}_{13}+i\chi^{\rm{re}}_{02} & \chi^{\rm{re}}_{23}-i\chi^{\rm{re}}_{01} & \chi^{\rm{re}}_{33} 
\end{pmatrix}.
\end{equation}
Let us rewrite it in other variables
\begin{equation}
\chi=
\begin{pmatrix} 
1-\chi_1-\chi_2-\chi_3 & \chi_4+i\chi_5 & \chi_6+i\chi_7 & \chi_8+i\chi_9\\
\chi_4-i\chi_5 & \chi_1 & \chi_{10}+i\chi_8 & \chi_{11}-i\chi_6\\
\chi_6-i\chi_7 & \chi_{10}-i\chi_8 & \chi_2 & \chi_{12}+i\chi_4\\
\chi_8-i\chi_9 & \chi_{11}+i\chi_6 & \chi_{12}-i\chi_4 & \chi_3 
\end{pmatrix}.
\end{equation}

In general, an arbitrary density matrix can be written as
$
\rho=
\left.\begin{pmatrix} 
1+P_z & P_x-iP_y \\ 
P_x+iP_y & 1-P_z
\end{pmatrix}\right/2.
$
Then measure of decoherence depends on the components of $ \chi $-matrix as follows:
\begin{multline}\label{eq:D_general}
D(\chi)=\max_{\substack{P_x,P_y,P_z}}\left| 2\chi_4 P_x+2\chi_6 P_y+2\chi_8 P_z \pm\vphantom{P_z^{1/2}}\right.\\
\pm \left[\left(\chi_{10}^2+\chi_{11}^2+\chi_2^2+\chi_3^2+\chi_7^2+\chi_9^2+2\chi_2\chi_3-2\chi_{11}\chi_7+2\chi_{10}\chi_9\right) 
P_x^2+ \right.\\
+ \left(\chi_{10}^2+\chi_{12}^2+\chi_1^2+\chi_3^2+\chi_5^2+\chi_9^2+2\chi_1\chi_3-2\chi_{10}\chi_9+2\chi_{12}\chi_5\right)
P_y^2+ \\
+ \left(\chi_{11}^2+\chi_{12}^2+\chi_1^2+\chi_2^2+\chi_5^2+\chi_7^2+2\chi_1\chi_2-2\chi_{12}\chi_5+2\chi_{11}\chi_7\right)
P_z^2+ \\
+ 2\left(2\chi_{10}\chi_3+\chi_{10}\chi_2+\chi_{12}\chi_7+\chi_5\chi_7+\chi_{10}\chi_1+\chi_9\chi_1-\chi_{11}\chi_{12}-
\chi_{11}\chi_5-\chi_2\chi_9\right)P_xP_y+ \\
+ 2\left(2\chi_{12}\chi_1+\chi_{12}\chi_2+\chi_{12}\chi_3+\chi_2\chi_5+\chi_{11}\chi_9+\chi_7\chi_9-\chi_{10}\chi_{11}-
\chi_{10}\chi_7-\chi_3\chi_5\right)P_yP_z+ \\
\left.\left.+ 2\left(2\chi_{11}\chi_2+\chi_{11}\chi_3+\chi_{10}\chi_5+\chi_1\chi_{11}+\chi_3\chi_7+\chi_5\chi_9-\chi_{10}\chi_{12}-
\chi_{12}\chi_9-\chi_1\chi_7\right)P_xP_z\right]^{1/2} \right|.
\end{multline}

In the case $ \chi_4=\chi_6=\chi_8=0$, the square of the measure of decoherence is a quadratic form, which can be reduced to canonical form by introducing the new variables $ Q_x,Q_y,Q_z $:
\begin{equation}
D^2=\max_{\substack{Q_x,Q_y,Q_z}}\left[a_1 Q_x^2+ a_2 Q_y^2+ a_3 Q_z^2\right].
\end{equation}

Let us consider the function $ f=a_1 Q_x^2+ a_2 Q_y^2+ a_3 Q_z^2 $, where the coefficients $ \{a_1, a_2, a_3\} $ are the eigenvalues of the quadratic form $ f $. With this change of variables:
\begin{equation}\label{eq:field}
\left\{
\begin{array}{rcl}
Q_x^2+ Q_y^2+ Q_z^2 \leq 1\\
Q_x^2 \geq 0, Q_y^2 \geq 0, Q_z^2 \geq 0\\
f \geq 0.\\
\end{array}
\right.
\end{equation}

Since for all $ Q_x, Q_y, Q_z $ the function $ f $ should be nonnegative, we get $ a_1 \geq 0, a_2 \geq 0, a_3 \geq 0  $. Hence the function $ f $ is nondecreasing, the maximum of this function is achieved on the boundary of Eq.~(\ref{eq:field}). The point $ (Q_x, Q_y, Q_z) $ of space at which there is the maximum of measure of decoherence is limited to the set $ \{(1,0,0), (0,1,0), (0,0,1)\} $. Therefore, we have
\begin{equation}\label{eq:specf}
D=\sqrt[]{\max\{\mathrm{spec}(f)\}}.
\end{equation} 

%%%%%%%%%%%%%%%%%%%%%%%%%%%%%%%%%%%%%%%%%%%%%%%%%%%%%%%%%%%%%
\section{SIMULATION OF NOISY QUANTUM CHANNELS AND QEC}

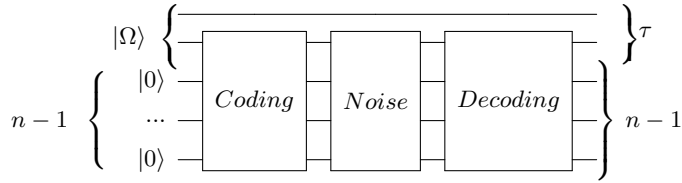
\begin{figure}
\begin{align*}
\Qcircuit @C=1em @R=.7em {
    \lstick{}  & \lstick{}  & \dstick{\ket{\Omega}}& \lstick{} & \lstick{}  & \qw & \qw & \qw  & \qw & \lstick{} & \dstick{\tau} \\ 
  \lstick{}  & \lstick{}  & \lstick{}  & \lstick{} & \lstick{} & \multigate{3}{Coding} & \multigate{3}{Noise} & \multigate{3}{Decoding} & \lstick{} \qw & \lstick{} & \lstick{} & \lstick{} & \lstick{}\\
  \lstick{}  & \lstick{}  & \lstick{}  & \lstick{}  & \lstick{\ket{0}}    & \ghost{Coding}      & \ghost{Noise} & \ghost{Decoding}        & \qw & \lstick{} & \lstick{} & \lstick{} & \lstick{}\\
 \lstick{n-1}  & \lstick{}  & \lstick{}  &  \lstick{}  & \lstick{...}    & \ghost{Coding}      & \ghost{Noise} & \ghost{Decoding}        & \qw & \lstick{} & \lstick{} & \lstick{} & \lstick{n-1}\\
  \lstick{}  & \lstick{}  & \lstick{}  & \lstick{}  & \lstick{\ket{0}}    & \ghost{Coding}    & \ghost{Noise} & \ghost{Decoding}        & \qw & \lstick{} & \lstick{} & \lstick{} & \lstick{}
  \gategroup{3}{2}{5}{3}{.7em}{\{} \gategroup{3}{7}{5}{9}{.7em}{\}}
  \gategroup{1}{5}{2}{6}{.7em}{\{} \gategroup{1}{9}{2}{10}{.7em}{\}}
}
\end{align*}
\caption{An n-qubit QECC and the scheme for calculating $ \tau $-matrix in case of arbitrary error correction. n-qubit QECC is shown on this Figure. $ \ket{\Omega} $ is a maximum entangled two-qubit pure state. The first qubit of the state $ \ket{\Omega} $ remains unchanged during the process of QEC, the second qubit of $ \ket{\Omega} $ is the first qubit of n-qubit QECC.}
\label{fig:general_scheme}
\end{figure}

To find an analytical expression for the measure of decoherence it is necessary to obtain $ \chi $-matrix, that describes the influence of an environment on the qubit. The given $ \chi $-matrix depends on parameter $ q $, it is the probability of the qubit to decohere due to interaction with an environment. The function $ D(p) $ can be calculated using Eqs.~(\ref{eq:D_diag_chi}),~(\ref{eq:D_general}) and~(\ref{eq:specf}). Despite the usability of $ \chi (p) $, the calculation of this function according to definition of $ \chi $-matrix in complicated. It is difficult to write a quantum operation in the form of Eq.~(\ref{eq:operator_summ}) in case of simulating QECC. This problem can be avoided by calculating Choi-Jamiołkowski state $ \tau $. The scheme for calculating $ \tau $-matrix in the case of QEC is shown on the Fig.~\ref{fig:general_scheme}. It is easy to find $ \hat{\hat{\chi}} $ using $ \tau $. From Eq.~(\ref{eq:choi_jam}) we can derive $ \chi_{\alpha\beta} $ coefficients which form $ \chi $-matrix. 

\begin{figure}
\begin{align*}
 \Qcircuit @C=1em @R=.7em {
 \dstick{\ket{\Omega}} & \lstick{} & \lstick{} & \qw & \qw & \lstick{} & \dstick{\tau} \\
 \dstick{} & \lstick{} & \lstick{} & \gate{Noise}   & \qw & \lstick{} \gategroup{1}{3}{2}{4}{.7em}{\{} \gategroup{1}{4}{2}{5}{.7em}{\}}
 }
\end{align*}
\caption{Scheme for calculating measure of decoherence in case of an error in a qubit. The first qubit remains unchanged, the second is affected by a quantum channel.}
\label{fig:qubit_scheme}
\end{figure}
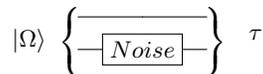

The above procedure can be summarized as the following sequence:
\begin{enumerate}
\item We write down the evolution of the density matrix for a single qubit interacted with an environment as an operator-sum. This decomposition depends on the parameter $ q $. The parameter characterizes the amount of errors.
\item We calculate $ \tau(q) $ state after the influence of an environment (Fig.~\ref{fig:qubit_scheme}). We obtain $ \hat{\hat{\chi}}(q) $ state using $ \tau(q) $.
\item From $ \hat{\hat{\chi}}(q) $ we get $ \chi(q) $-matrix.
\item We calculate measure of decoherence $ D_0(q) $ for qubit in an environment. Here $ D_0(q) $ is a probability for $ p $ to decohere.
\item We replace $ q $ with $ p $ in the operator-sum based on the equality $ D_0(q) \equiv p $.
\item We obtain the limitation on $ p $ based on the positivity of $ \chi(p) $-matrix.
\item Next we consider the error correction code. We take $ N $ different values of $ p $, using the existing limitations on the probability $ p $. We calculate Choi-Jamiołkowski state $ \tau(p) $ after applying QECC. We get $ \hat{\hat{\chi}}(p) $ state using $ \tau(p) $.
\item From $ \hat{\hat{\chi}}(p) $ we get $ \chi(p) $-matrix.
\item We calculate measure of decoherence $ D(p) $ of qubit after interaction with an environment and quantum correction. For $ n $ different $ p $ we take $ n $ values of $ D $, $ n $ is the number of qubits in QECC.
\item Since the quantum correction code gives a polynomial in $ p $ improvement we have
\begin{equation}
D(p)=\sum_{i=1}^n \alpha_i p^i.
\end{equation}
We solve $ n $ equations and calculate coefficients $ \{\alpha_i\}_{i=1}^n $ of the polynomial of the $ n $-th degree. Hence we establish polynomial expression for $ D $.
\end{enumerate}

Next we consider the results of the algorithm for calculating $ D(p) $.

%%-----------------------------------------------------------
\subsection{Bit and Phase Flips}

Let bit flip environment changes $ \ket{0} $ to $ \ket{1} $, and $ \ket{1} $ to $ \ket{0} $ with probability $ p' $. This action can be conveniently written as Eq.~(\ref{eq:operator_summ}) with the Pauli matrices $\{\hat{I},\hat{X},\hat{Y},\hat{Z}\}$ as the basis $\{\hat{E}_{\alpha}\}_{\alpha=0}^{d^2-1}$ for $ d=2 $ (single qubit). Then the operator-sum is written as:
\begin{equation}\label{eq:bit_kraus_summ}
\mathcal{E}[\rho]=(1-p')\hat{I}\rho \hat{I}+p'\hat{X}\rho \hat{X}.
\end{equation}
This representation in the form of Kraus decomposition allows us to write $ \chi $-matrix of the quantum channel as a diagonal matrix in Eq.~(\ref{eq:chi_diag}) with parameters $ \chi_0 = 1-p' $, $ \chi_1 = p' $ and $ \chi_3 = \chi_4 = 0 $. It follows from Eq.~(\ref{eq:D_diag_chi}) that measure of decoherence $ D_0 \equiv p = p' $. Then $ \chi $-matrix of the quantum channel in case of bit flip is equal to
\begin{equation}\label{eq:chi_bit}
\chi_0=\begin{pmatrix} 1 - p & 0 & 0 & 0\\ 0 & p & 0 & 0\\ 0 & 0 & 0 & 0\\ 0 & 0 & 0 & 0\end{pmatrix}.
\end{equation}
According to $ \chi $-matrix positivity criterion, bit flip channel is completely positive if and only if $0 \le p\le 1$.

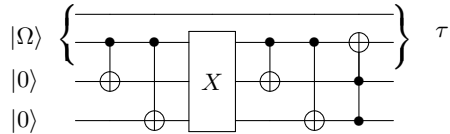
\begin{figure}
\begin{align*}
 \Qcircuit @C=1em @R=.7em {
 \dstick{\ket{\Omega}}& \lstick{}  & \lstick{} & \qw & \qw & \qw & \qw & \qw & \qw & \qw  & \lstick{} & \dstick{\tau} \\
 \dstick{} & \lstick{} & \lstick{} & \ctrl{1} & \ctrl{2} & \multigate{2}{X} & \ctrl{1} & \ctrl{2} & \targ & \qw & \lstick{} & \dstick{} \\
 \dstick{} & \lstick{\ket{0}} & \lstick{}  & \targ    & \qw      & \ghost{X}        & \targ    & \qw      & \ctrl{-1} & \qw \\
 \dstick{} & \lstick{\ket{0}} & \lstick{}  & \qw      & \targ    & \ghost{X}        & \qw      & \targ    & \ctrl{-2} & \qw
 \gategroup{1}{3}{2}{4}{.7em}{\{} \gategroup{1}{9}{2}{10}{.7em}{\}}
 }
\end{align*}
\caption{Scheme for calculating measure of decoherence in case of bit flip correction.}
\label{fig:bit_scheme}
\end{figure}

Let us consider a bit error correction code based on a decoding of type called majority voting~\cite{nielsen2010quantum}. Scheme for finding measure of decoherence in the case of majority voting is represented in Fig.~\ref{fig:bit_scheme}.

$ \chi $-matrix is calculated following the above algorithm:
\begin{equation}\label{eq:chi_bit_QECC}
\chi=\begin{pmatrix} 1 - 3 p^2 + 2 p^3 & 0 & 0 & 0\\ 0 & 3 p^2 - 2 p^3 & 0 & 0\\ 0 & 0 & 0 & 0\\ 0 & 0 & 0 & 0\end{pmatrix}.
\end{equation}
According to Eq.~(\ref{eq:D_diag_chi}), the measure of decoherence 
\begin{equation}
D(p)=3 p^2 - 2 p^3.
\end{equation}

The procedure of finding the result of an action and the result of a correction in case of phase flip is similar to the case of bit error correction. The only difference is the modified scheme of majority voting, shown in Fig.~\ref{fig:phase_scheme}.
\begin{figure}
\begin{align*}
 \Qcircuit @C=1em @R=.7em {
 \dstick{\ket{\Omega}}& \lstick{}  & \lstick{} & \qw & \qw & \qw & \qw & \qw & \qw & \qw & \qw & \qw  & \lstick{} & \dstick{\tau} \\
 \dstick{} & \lstick{} & \lstick{} & \ctrl{1} & \ctrl{2} & \gate{H} & \multigate{2}{Z} & \gate{H} & \ctrl{1} & \ctrl{2} & \targ & \qw & \lstick{} & \dstick{} \\
 \dstick{} & \lstick{\ket{0}} & \lstick{}  & \targ    & \qw  & \gate{H}    & \ghost{X}   & \gate{H}     & \targ    & \qw      & \ctrl{-1} & \qw \\
 \dstick{} & \lstick{\ket{0}} & \lstick{}  & \qw      & \targ  & \gate{H}  & \ghost{X}    & \gate{H}    & \qw      & \targ    & \ctrl{-2} & \qw
 \gategroup{1}{3}{2}{4}{.7em}{\{} \gategroup{1}{11}{2}{12}{.7em}{\}}
 }
\end{align*}
\caption{Scheme for calculating measure of decoherence in case of phase flip correction.}
\label{fig:phase_scheme}
\end{figure}

%%-----------------------------------------------------------
\subsection{Depolarizing Channel}
It is convenient to write an operator sum in case of the depolarizing channel in the form of Kraus decomposition:
\begin{equation}\label{eq:depolar_kraus_summ}
\mathcal{E}[\rho]=\left(1-\dfrac{3}{2}p\right)\hat{I}\rho \hat{I}+\dfrac{p}{2}\left(\hat{X}\rho \hat{X}+\hat{Y}\rho \hat{Y}+\hat{Z}\rho \hat{Z}\right).
\end{equation}
Since all the matrices in the decomposition are the Pauli matrices, the matrix $  \chi_0 $ is diagonal and is equal to
\begin{equation}
\chi_0=\begin{pmatrix} 1-\dfrac{3}{2}p & 0 & 0 & 0\\ 0 & p/2 & 0 & 0\\ 0 & 0 & p/2 & 0\\ 0 & 0 & 0 & p/2\end{pmatrix}.
\end{equation}
The parameters in the decomposition in Eq.~\ref{eq:depolar_kraus_summ} are selected to satisfy $ D_0 = p $. From positivity of $ \chi_0 $-matrix $ p \in [0,2/3] $.

\begin{figure}
\begin{align*}
 \Qcircuit @C=1em @R=.7em {
   \dstick{\ket{\Omega}}& \lstick{}   & \lstick{} & \qw & \qw & \qw & \qw & \qw & \qw & \qw & \qw & \qw & \qw & \qw & \lstick{} & \dstick{\tau} \\ 
  \dstick{}& \lstick{}  & \lstick{} & \gate{Z} & \targ                 & \qw       & \targ \qwx[1]       & \qw 		& \gate{Z} \qwx[2] & \qw & \qw 				      & \multigate{4}{E_{\text{all}}}      & \multigate{4}{R}          & \lstick{} \qw \\
  \dstick{} & \lstick{\ket{0}}  & \lstick{} & \gate{H} & \ctrl{-1} \qwx[1] & \qw       & \gate{Z} \qwx[2] & \qw 		& \qw 					  & \qw & \gate{Z} \qwx[1] & \ghost{E_{\text{all}}}                & \ghost{R}                    & \qw \\
  \dstick{} & \lstick{\ket{0}}  & \lstick{} & \qw     & \gate{Z} \qwx[2] & \qw       & \qw                       & \qw 		& \gate{Z} \qwx[1] & \gate{H} & \ctrl{2} 		   & \ghost{E_{\text{all}}}                & \ghost{R}                    & \qw \\
  \dstick{} & \lstick{\ket{0}}  & \lstick{} & \qw       & \qw                   & \qw       & \gate{Z}                & \gate{H}  & \ctrl{1}				  & \qw & \targ 				   & \ghost{E_{\text{all}}}                 & \ghost{R}                   & \qw \\
  \dstick{} & \lstick{\ket{0}}  & \lstick{} & \qw       & \gate{Z}             & \gate{H} & \ctrl{-1}               & \qw 		& \targ 				  & \qw & \gate{Z} 		    	& \ghost{E_{\text{all}}}                 & \ghost{R}                   & \qw
    \gategroup{1}{3}{2}{4}{.7em}{\{} \gategroup{1}{13}{2}{14}{.7em}{\}}
}
\end{align*}
\caption{5-qubit DiVincenzo-Shor QECC. R is the unitary operator~\cite{tomita2011unitary, melnikov2013quantum}. The code does not contain any measurement operations.}
\label{fig:5QEC_scheme}
\end{figure}

Let us consider the result of the 5-qubit DiVincenzo-Shor QEC~\cite{divincenzo1996fault, tomita2011unitary}, shown in Fig.~\ref{fig:5QEC_scheme}, after the action of the depolarizing channel. In the case of the depolarizing channel it is possible to obtain a general view of $ \chi $-matrix, which describes the action of the error correction code:
\begin{equation}\label{eq:chi_DepChannel_5QECC}
\chi=\begin{pmatrix}
1-3\chi_{1} & 0 & 0 & 0\\ 0 & \chi_{1} & 0 & 0\\
0 & 0 & \chi_{1} & 0\\ 0 & 0 & 0 & \chi_{1}
\end{pmatrix},
\end{equation}
where
\begin{equation}
\chi_{1}=p^2 \left.\left(15 - 50 p + 60 p^2 - 24 p^3\right)\right/2.
\end{equation}
It follows from Eq.~(\ref{eq:D_diag_chi}) that measure of decoherence is equal to 
\begin{equation}\label{eq:DepolarDiV}
D(p)=p^2 \left(15 - 50 p + 60 p^2 - 24 p^3\right).
\end{equation}

\begin{figure}
\begin{align*}
 \Qcircuit @C=1em @R=.7em {
   \dstick{\ket{\Omega}}& \lstick{}   & \lstick{} & \qw & \qw & \qw & \qw & \qw & \qw & \qw & \qw & \qw & \qw & \qw & \qw & \qw  & \qw  & \lstick{} & \dstick{\tau} \\
  \dstick{}& \lstick{}  & \lstick{}  & \ctrl{6} & \ctrl{3} & \gate{H} & \ctrl{1} & \ctrl{2} & \multigate{8}{E_{\text{all}}} & \ctrl{1} & \ctrl{2} & \targ & \gate{H} & \ctrl{6} & \ctrl{3} & \targ  &  \qw & \lstick{} \\
  \dstick{} & \lstick{\ket{0}}  & \lstick{}    & \qw  & \qw  & \qw  & \targ   &  \qw & \ghost{E_{\text{all}}}  & \targ    & \qw  & \ctrl{-1} &  \qw & \qw  & \qw & \qw & \qw \\
  \dstick{} & \lstick{\ket{0}}  & \lstick{}    & \qw      & \qw    &  \qw & \qw      & \targ  & \ghost{E_{\text{all}}}  & \qw      & \targ    & \ctrl{-2} &  \qw & \qw & \qw  & \qw  & \qw \\
 \dstick{} & \lstick{\ket{0}}  & \lstick{}    & \qw     &  \targ    & \gate{H} & \ctrl{1} & \ctrl{2}  & \ghost{E_{\text{all}}} & \ctrl{1} & \ctrl{2} & \targ & \gate{H}  & \qw     &  \targ   & \ctrl{-3} &   \qw \\
  \dstick{} & \lstick{\ket{0}}  & \lstick{}    & \qw     & \qw    &  \qw & \targ   &  \qw  & \ghost{E_{\text{all}}}  & \targ    & \qw  & \ctrl{-1} &  \qw  & \qw    & \qw  & \qw & \qw \\ 
  \dstick{} & \lstick{\ket{0}} & \lstick{}    & \qw     & \qw    &  \qw & \qw      & \targ  & \ghost{E_{\text{all}}}  & \qw      & \targ    & \ctrl{-2} &  \qw  & \qw    & \qw  & \qw & \qw \\ 
  \dstick{} & \lstick{\ket{0}}  & \lstick{}    & \targ     & \qw    & \gate{H} & \ctrl{1} & \ctrl{2}  & \ghost{E_{\text{all}}}  & \ctrl{1} & \ctrl{2} & \targ & \gate{H}  & \targ     & \qw   & \ctrl{-6} & \qw \\ 
  \dstick{} & \lstick{\ket{0}} & \lstick{}    & \qw     & \qw    &  \qw & \targ   &  \qw  & \ghost{E_{\text{all}}} & \targ    & \qw  & \ctrl{-1} &  \qw  & \qw    & \qw  & \qw & \qw \\
  \dstick{} & \lstick{\ket{0}} & \lstick{}    & \qw     & \qw    &  \qw & \qw      & \targ  & \ghost{E_{\text{all}}} & \qw      & \targ    & \ctrl{-2} &  \qw  & \qw    & \qw  & \qw & \qw
  \gategroup{1}{3}{2}{4}{.7em}{\{} \gategroup{1}{16}{2}{17}{.7em}{\}}
}
\end{align*}
\caption{9-qubit Shor QECC. This code is a composition of bit and phase flip error correction codes shown in Figs.~\ref{fig:bit_scheme} and~\ref{fig:phase_scheme}. The code does not contain any measurement operations.}
\label{fig:9QEC_scheme}
\end{figure}
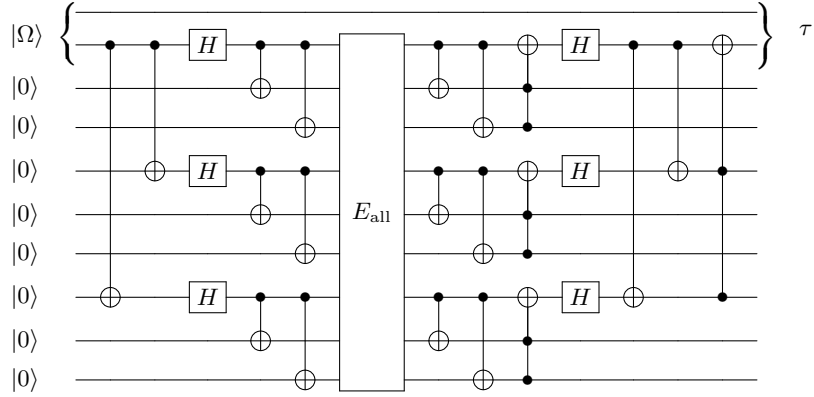

Let $ p\to 0 $. For 9-qubit Shor code~\cite{shor1995scheme} shown in Fig.~\ref{fig:9QEC_scheme} measure of decoherence has the form
\begin{equation}\label{eq:DepolarShor}
D(p)=36p^2.
\end{equation}

\begin{figure}
   \begin{center}
   \begin{tabular}{c}
   \includegraphics[height=7cm]{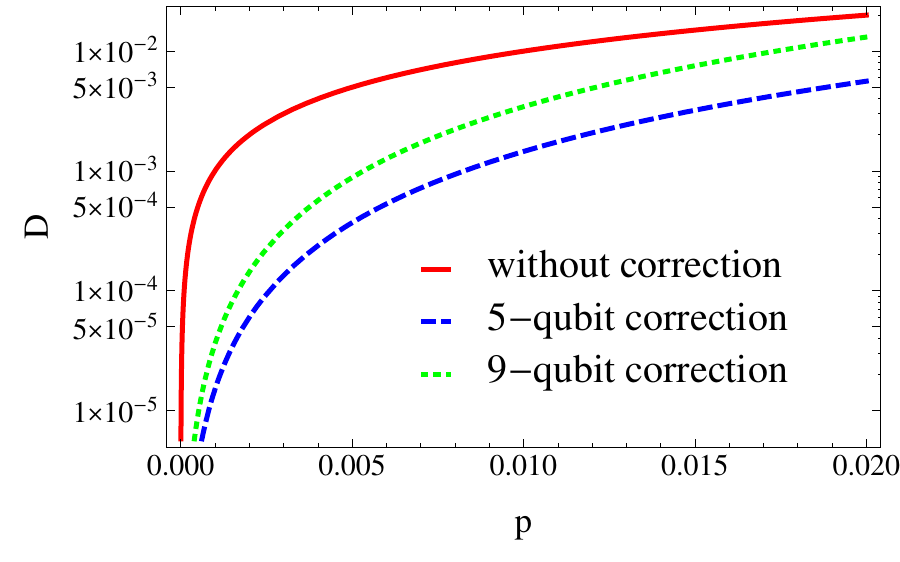}
   \end{tabular}
   \end{center}
\caption{QEC in case of a depolarizing channel. Measure of decoherence D vs. the probability of an error in a qubit p. The results are consistent with the numerical simulation~\cite{melnikov2013quantum}.}
\label{fig:QECresultDepolar}
   \end{figure}

Combining Eqs.~(\ref{eq:DepolarDiV}) and~(\ref{eq:DepolarShor}), we obtain the result of QEC shown in Fig.~\ref{fig:QECresultDepolar} in case of the depolarizing channel.

%-------------
%-------------
 
 %%%%%%%%%%%%%%%%%%%%%%%%%%%%%%%%%%%%%%%%%%%%%%%%%%%%%%%%%%%%%
\section{QEC IN SI DOUBLE DOT CHARGE QUBITS}

\subsection{Amplitude Damping}\label{subsec:DQDAmpl}
Let us consider the basis $\ket{+}=\frac{1}{\sqrt{2}}[\ket{0}+\ket{1}]$, $\ket{+}=\frac{1}{\sqrt{2}}[\ket{0}-\ket{1}]$. In this basis the evolution of the density matrix is written in the form of
\begin{equation}\label{eq:DGDEvolutionA}
\begin{pmatrix} 1-\rho_{--}(0)e^{-\Gamma  t} & \rho_{+-}(0)e^{-\Gamma t /2} \\
\rho_{-+}(0)e^{-\Gamma t /2} & \rho_{--}(0)e^{-\Gamma t} \end{pmatrix},
\end{equation}
denote by $\Gamma(t)$ a relaxation rate. Let us rewrite Eq.~(\ref{eq:DGDEvolutionA}) to obtain the representation in the form of an operator-sum:
\begin{equation}\label{eq:operator_dqdA}
\mathcal{E}[\rho]=\begin{pmatrix} 1 & 0 \\ 0 & e^{-\Gamma t /2} \end{pmatrix}\rho \begin{pmatrix} 1 & 0 \\ 0 & e^{-\Gamma t /2} \end{pmatrix}^\dag+
\begin{pmatrix} 0 & \sqrt{1-e^{-\Gamma t}} \\ 0 & 0 \end{pmatrix}\rho \begin{pmatrix} 0 & \sqrt{1-e^{-\Gamma t}} \\ 0 & 0 \end{pmatrix}^\dag.
\end{equation}

Let us consider the state $\tau$. The nonzero eigenvalues of the state are $ (1\pm e^{-\Gamma t})/2 $. Since $\Gamma t\ge 0$, we see that all eigenvalues of the matrix $\tau$ are nonnegative, that is $\tau$ is positive semidefinite. Therefore, the map in Eq.~(\ref{eq:operator_dqdA}) is completely positive.

We represent the quantum operation shown in Eq.~(\ref{eq:operator_dqdA}) in the form of $ \chi_0 $-matrix in Pauli matrix basis:
\begin{equation}\label{eq:chi_gamma}
\chi_0=\left.\begin{pmatrix} \left(1+e^{-\Gamma t/2}\right)^2 & 0 & 0 & (1-e^{-\Gamma t})\\ 0 & (e^{-\Gamma t}-1) & i(e^{-\Gamma t}-1) & 0\\ 0 & -i(e^{-\Gamma t}-1) & (e^{-\Gamma t}-1) & 0\\ (1-e^{-\Gamma t}) & 0 & 0 & \left(1-e^{-\Gamma t/2}\right)^2\end{pmatrix}\right/4.
\end{equation}
According to Eq.~(\ref{eq:measureDdefinition}) measure of decoherence is equal to $ D_0 = 1-e^{\Gamma t} = p$. It follows from the proposed algorithm that measure of decoherence in case of the 5-qubit QECC
\begin{equation}\label{eq:measureDampDamp}
D(p) = 5p^2\left.\left(3 - 3p + p^2\right)\right/8.
\end{equation}
It can be concluded that for small $ p $ the efficiency of the error correction in the case of amplitude damping is 8 times higher than in the case of the depolarizing channel.

\subsection{Phase Damping}\label{subsec:DQDPhase}
The evolution of the density matrix in a single operation can be written as
\begin{equation}\label{eq:DGDEvolutionP}
\begin{pmatrix} \rho_{00}(0) & \rho_{01}(0)e^{-B^{2}} \\
\rho_{10}(0)e^{-B^{2}} & \rho_{11}(0) \end{pmatrix},
\end{equation}
denote by $B^2(t)$ a spectral function. Using representation in Eq.~(\ref{eq:DGDEvolutionP}), we can write the operator-sum:
\begin{multline}\label{eq:operator_dqdAB}
\mathcal{E}[\rho]=\begin{pmatrix} e^{-B^{2}/2} & 0 \\ 0 & e^{-B^{2}/2} \end{pmatrix}\rho \begin{pmatrix} e^{-B^{2}/2} & 0 \\ 0 & e^{-B^{2}/2} \end{pmatrix}^\dag +\\
+\begin{pmatrix} \sqrt{1-e^{-B^{2}}} & 0 \\ 0 & 0 \end{pmatrix}\rho \begin{pmatrix} \sqrt{1-e^{-B^{2}}} & 0 \\ 0 & 0 \end{pmatrix}^\dag+\begin{pmatrix} 0 & 0 \\ 0 & \sqrt{1-e^{-B^{2}}} \end{pmatrix}\rho \begin{pmatrix} 0 & 0 \\ 0 & \sqrt{1-e^{-B^{2}}} \end{pmatrix}^\dag.
\end{multline}

Consider Choi-Jamiołkowski state $\tau$. The nonzero eigenvalues of the state are $ \left.\left(1\pm e^{-B^{2}}\right)\right/2 $. Since $B^{2}\ge 0$, we have nonnegative eigenvalues of matrix $\tau$, that is the state $\tau$ is positive semidefinite. Hence, the map in Eq.~(\ref{eq:operator_dqdAB}) is completely positive.

Let us represent the quantum operation of Eq.~(\ref{eq:operator_dqdAB}) in the form of $ \chi_0 $-matrix in Pauli matrix basis:
\begin{equation}
\chi_0=\begin{pmatrix} \left(1+e^{-B^{2}}\right)/2 & 0 & 0 & 0\\ 0 & 0 & 0 & 0\\ 0 & 0 & 0 & 0\\ 0 & 0 & 0 & \left(1-e^{-B^{2}}\right)/2\end{pmatrix}.
\end{equation}
It follows from Eq.~(\ref{eq:D_diag_chi}) that measure of decoherence is equal to $ D_0=\left.\left(1-e^{-B^{2}}\right)\right/2 = p$. According to the proposed algorithm we calculate measure of decoherence in case of the 5-qubit QECC
\begin{equation}\label{eq:measureDphaseDamp}
D(p) = 10p^2(1 - 2p + p^2).
\end{equation}
It can be concluded that for small $ p $ the efficiency of the error correction in the case of amplitude damping is 1.5 times higher than in the case of the depolarizing channel.

\subsection{QEC Results}
During calculations, transmission or information storage there occurs amplitude and phase damping. To fight the process of decoherence one can use error correction algorithms, which have already been discussed in this paper.

Let us consider an error correction in silicon, where the qubit represents as a state of an electron in the double quantum dot~\cite{fedichkin2000coherent, fedichkin2004error, fedichkin2005study, filippov2011effect}. To describe the entire system of five qubit we use our results for measure of decoherence in Eqs.~(\ref{eq:measureDampDamp}) and~(\ref{eq:measureDphaseDamp}). To determine relaxation rate and spectral function, we use the following formulas
\begin{equation}
\Gamma=\frac{\Xi^2k^3}{4\pi\rho s^2\hbar}\exp(-a^2k^2/2)\left(1-\frac{\sin(kL)}{kL}\right),
\end{equation}
\begin{equation}
B^2(t)=\frac{\Xi^2}{\pi^2\hbar\rho s^3}\int\limits_0^\infty q^2dq\int\limits_0^\pi\sin\Theta d\Theta \frac{\sin^2(qL\cos\Theta)\exp(-a^2q^2/2)}{q}\sin^2\frac{qst}{2},
\end{equation}
where deformation potential $\Xi = 3.3 $ eV, speed of sound $s = 9.0\cdot 10^3$ m/s, crystal density $\rho = 2.33$ g/sm$^3$, the distance between points $L = 50$ nm, radius of points $a = 3$ nm.

%-------------
   \begin{figure}
   \begin{center}
   \begin{tabular}{c}
   \includegraphics[height=7cm]{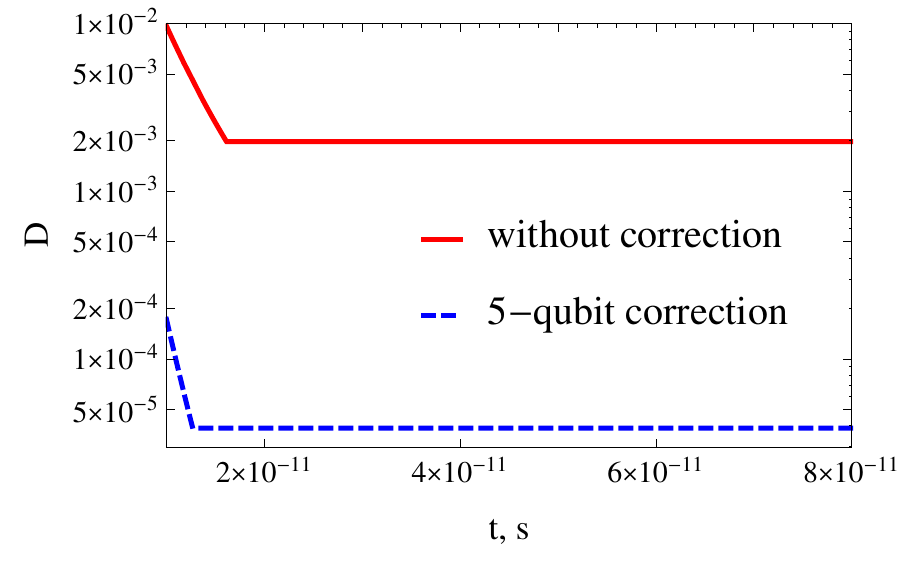}
   \end{tabular}
   \end{center}
\caption{QEC in Si double dot charge qubit. Measure of decoherence D vs. the computational cycle time t. The figure shows the error correction after N=68 operations. The result are consistent with the numerical simulation~\cite{melnikov2013quantum}.}
\label{fig:QECresultDQD}
   \end{figure} 
%-------------

Depending upon the quantum gate operation required at current algorithmic step, a qubit experience phase or amplitude damping error during quantum computing on double dot qubit. Hence we define measure of decoherence in this case as the maximum of the two measures. Combining Eqs.~(\ref{eq:measureDampDamp}) and~(\ref{eq:measureDphaseDamp}), we obtain
\begin{equation}\label{eq:measureDdot}
D_0(t) = \max \{p_1, p_2\},
\end{equation}
\begin{equation}\label{eq:measureDdotCorrection}
D(t) = \max \left\{~5p_1^2(3 - 3p_1 + p_1^2)/8~,~10p_2^2(1 - 2p_2 + p_2^2)~\right\},
\end{equation}
where $p_1 = 1-e^{\Gamma (t) t}$ and $p_2 = \left.\left(1-e^{-B(t)^{2}}\right)\right/2$.

Let quantum computer perform $ N $ operations. By $ p $ denote the probability of an error during one cycle time. Substituting $ p_1 $ for $ N p_1 $ and $ p_2 $ for $ N p_2 $ in Eqs.~(\ref{eq:measureDdot}) and~(\ref{eq:measureDdotCorrection}), we obtain the result of QEC shown in Fig.~\ref{fig:QECresultDQD}.

 %%%%%%%%%%%%%%%%%%%%%%%%%%%%%%%%%%%%%%%%%%%%%%%%%%%%%%%%%%%%%
\section{CONCLUSIONS}
By using $ \chi $-matrix, Choi-Jamiołkowski state, and measure of decoherence techniques we analysed the influence of noise on quantum bits. The analytical expressions of measure of decoherence were obtained for practically important wide subset of quantum channels. We have shown that the introduction of DiVincenzo-Shor quantum error correction algorithm is helpful for double dot charge qubits and reduces effective error rate drastically.

%%%%%%%%%%%%%%%%%%%%%%%%%%%%%%%%%%%%%%%%%%%%%%%%%%%%%%%%%%%%%
\section*{ACKNOWLEDGMENTS}
 
%A.A.M. thanks the Russian Foundation for Basic Research (Project No. 12-02-31524).
The work was supported via the grant No. 07.524.12.4019 of the Ministry of Education and Science of the Russian Federation.

%%%%%%%%%%%%%%%%%%%%%%%%%%%%%%%%%%%%%%%%%%%%%%%%%%%%%%%%%%%%%
%%%%% References %%%%%

\bibliography{report}   %>>>> bibliography data in report.bib
\bibliographystyle{spiebib}   %>>>> makes bibtex use spiebib.bst

\end{document}